\documentclass[11pt]{article}
\usepackage{hyperref}
\pdfoutput=1 

\begin{document} 

\title{Interactions of Shear Layer Vortices with the Trailing Corner \\in an Open Cavity Flow} 

\author{Xiaofeng Liu, Joseph Katz \\
\\\vspace{6pt} Department of Mechanical Engineering, \\ Johns Hopkins University, Baltimore, Maryland 21218, USA\\Email: XiaofengLiu@jhu.edu \\ http://www.me.jhu.edu/lefd/PPIV/} 

\maketitle 


\begin{abstract}
This fluid dynamics video provides sample experimental results focusing on the interactions of shear layer vortices with the trailing corner in a 2D open cavity shear layer. These interactions were investigated experimentally in a water tunnel at a Reynolds number of $4.0\times 10^4$.  Time-resolved particle image velocimetry (PIV) with an image sampling rate of 4500 frames per second was used to simultaneously measure the instantaneous velocity, material acceleration and pressure distributions. The latter was calculated by integrating the spatial distribution of in-plane components of the material acceleration. A large database of instantaneous realizations visualized the dynamic changes to the shear layer vortices, such as deformation and breakup as they impinged and climbed over the cavity trailing corner. These interactions cause time-dependent formation of a pressure maximum as the flow impinges on the forward facing surface of the trailing corner, and a minimum above the corner, where large local pressure gradients dominate the generation of fresh vorticity. Data analysis shows that interactions with the corner involve two time scales. Intermediate Strouhal numbers (0.6-3.2) are associated with interaction of the shear layer vortices with the corner, while vertical flapping of the shear layer occurs at low Strouhal numbers (0.1-0.6). 
\end{abstract} 


\section{Introduction} 

This fluid dynamics video demonstrates the results of \href{http://www.me.jhu.edu/lefd/PPIV/}{an experimental investigation} of the 
interactions of shear layer vortices with the trailing corner in a turbulent open cavity flow recently conducted at Johns Hopkins University. 
Time resolved 2D PIV measurements with an image-sampling rate of 4500 frames per second and a field of view of $25\times 25$ mm have been used for studying the unsteady flow structure in the cavity shear layer, with an emphasis on pressure-related flow phenomena that have not been thoroughly investigated due to the lack of experimental capabilities.

Details about the experimental setup can be found at Liu and Katz (2011a and 2011b). The 2-D cavity has a length ($L$) of 38.1 mm and a depth ($D$) of 30.0 mm, thus forming a moderate length to depth ratio of 1.27, ensuing an open cavity flow (shear layer impinges on the vertical trailing wall only).  
A 13 mm long region with tripping grooves is machined at the beginning of the contraction ramp in order to trip the boundary layer. Therefore, 
the separating boundary layer at the beginning of the cavity is turbulent.
The mean velocity above the cavity is ${U_\infty}$=1.20 m/s, corresponding to Reynolds numbers of $4.0\times 10^4$ based on cavity length. The origin of the coordinate system used in the video is placed at the leading edge of the cavity, with the x and y axes pointing downstream and upward, respectively. The instantaneous and the fluctuating values of the horizontal velocity, the vertical velocity and the pressure are denoted as $u$, $v$, $p$, $u'$ , $v'$, and $p'$, respectively. The swirling strength, $\lambda$, i.e., the imaginary part of the complex eigenvalue of the local velocity gradient tensor (Zhou $et$  $al$ 1996), represents the strength of local swirling motion and is used in this video to identify vortices in the shear layer.    

Simultaneously obtained instantaneous velocity, material acceleration and pressure distributions are available with the time-resolved measurements.  
The pressure was obtained by spatially integrating the material acceleration, following the procedures that we introduced a few years ago (Liu and Katz, 2006). This technique originally utilized four-exposure PIV to measure the distribution of the in-plane components of the material acceleration, 
and then integrating them by means of a virtual boundary, omni-directional integration algorithm to obtain the pressure distribution. 
The robustness of this integration method has been confirmed recently by Charonko $et$ $al$ (2010), and utilized by several groups. 
Using this technique, we measured the velocity and pressure distributions in a high Reynolds number open cavity shear layer flow, 
and obtained results that are in agreement with the physical location and appearance of cavitation as well as the independently-measured cavitation inception indices (Liu and Katz 2008). 
This experiment revealed that the lowest pressure, and consequently, the location of cavitation inception, developed periodically above the trailing corner of the cavity.  

The instantaneous pressure distribution is a key quantity associated with vortical motions in shear layer flows and their interactions among themselves and with solid boundaries.  Prediction of the pressure field is also essential for prediction of noise generation and flow-induced vibrations. In this video, we use sample data obtained from recent time-resolved PIV measurements to show how interactions of the cavity shear layer with the trailing corner affect the unsteady flow and pressure fluctuations around the corner.  The video contains an illustration of the experimental setup, a short excerpt of sample particle images, and three parts of sample movies showing the time-evolution of instantaneous distribution of selected flow quantities including swirling strength, pressure and shear strain rate. 
In the sample particle image movie, an intermittent small separation (reverse flow) region with a height of about 0.1 mm can be seen right on top of the trailing corner.  The pseudostreamlines superimposed on the distributions of flow quantities in the sample movies are obtained by subtracting half of the free stream velocity at every point. 
Sample I shows the dynamic changes to shear layer vortices, such as deformation and breakup as they impinge and climb over the cavity trailing corner when the shear layer is located at high elevation due to low-frequency undulation/flapping. The periodic appearance and disappearance of high and low pressure peaks in front of and above the trailing corner, synchronized with the streamwise position of the large vortical structures, is evident from the sample movie. As demonstrated in Liu and Katz (2011b), the downwash induced by the large eddies is responsible for the disappearance of the pressure minimum on top of the corner. In sample I, the vortex attached/latched to the top of the trailing corner is a result of intermittent local vorticity generation induced by high favorable pressure gradient around the corner. 
Sample II shows the behavior of the shear layer when it is located at low elevation during the low-frequency undulation cycle. In this case, substantial fraction of the shear layer is entrained back into the cavity. Recirculation of this flow back to the beginning of the cavity is part of the feedback mechanism causing the low frequency undulations of the shear layer. Sample III shows that the shear layer vortices are subjected to strong shear strain as they impinge on the trailing corner.

\section{Reference}

Adrian, R.J., Christensen, K.T. and Liu, Z.-C., 2000, "Analysis and interpretation of instantaneous turbulent velocity fields". Exp. Fluids. 29, 275-290.

Charonko, J. J., King, C. V., Smith, B. L. and Vlachos, P. P., 2010, "Assessment of pressure field calculations from particle image velocimetry measurements". Meas. Sci. and Tech., 21, 105401.

Liu, X. and Katz, J., 2006, "Instantaneous pressure and material acceleration measurements using a four exposure PIV system". Exp Fluids, 41,227-240.

Liu, X. and Katz, J., 2007, "A comparison of cavitation inception index measurements to the spatial pressure distribution within a 2D cavity shear flow". FEDFM2007-37090. 

Liu, X. and Katz, J., 2008, "Cavitation phenomena occurring due to interaction of shear layer vortices with the trailing corner of a 2D open cavity". Phys. of Fluids, 20, 041702.

Liu, X. and Katz, J., 2011a, "Time resolved PIV measurements elucidate the feedback mechanism that causes low-frequency undulation in an open cavity shear layer". 9th International Symposium on Particle Image Velocimetry - PIV11, Kobe, Japan, July 21-23, 2011.

Liu, X. and Katz, J. 2011b, "Time resolved measurements of the pressure field generated by vortex-corner interactions in a cavity shear layer". Proceedings of the ASME-JSME-KSME joint fluids engineering conference 2011, 
AJK2011-FED, Hamamatsu, Shizuoka, Japan, July 24-29. AJK2011-08018.

Gopalan, S. and Katz, J., 2000, "Flow structure and modeling issues in the closure region of
cavitation". Physics of Fluids, 12, 895-911.

Zhou, J., Adrian, R.J. and Balachandar, S., 1996, "Autogeneration of near-wall vortical structures in channel flow". Physics of Fluids. 8, 288-290.

Zhou, J., Adrian, R.J., Balachandar, S. and Kendall, T.M., 1999, "Mechanisms for generating coherent packets of hairpin vortices in channel flow". J. Fluid Mech. 387, 353-359.

\end{document}